\documentclass[letterpaper, 10 pt, conference]{ieeeconf}  % Comment this line out if you need a4paper

\IEEEoverridecommandlockouts                              % This command is only needed if 
\usepackage{hyperref}

\title{\LARGE \bf
``Who Should I Believe?": User Interpretation and Decision-Making When a Family Healthcare Robot Contradicts Human Memory
}

\author{Hong Wang$^{1}$, Natalia Calvo-Barajas$^{1}$, Katie Winkle$^{1}$, and Ginevra Castellano$^{1}$% <-this % stops a space
\thanks{*This work was partly supported by the Horizon Europe SymAware project (project number: 101070802); and the Graduate School in Cybersecurity, Department of Information Technology, Uppsala University}% <-this % stops a space
\thanks{$^{1}$Hong Wang, Natalia Calvo-Barajas, Katie Winkle, and Ginevra Castellano are with Uppsala University, Uppsala, Sweden.
        {\tt\small hong.wang@it.uu.se, katie.winkle@it.uu.se, natalia.calvo@it.uu.se, ginevra.castellano@it.uu.se}}%
}
\usepackage{graphicx}
\usepackage{multirow}
\begin{document}

\maketitle
\thispagestyle{empty}
\pagestyle{empty}

%%%%%%%%%%%%%%%%%%%%%%%%%%%%%%%%%%%%%%%%%%%%%%%%%%%%%%%%%%%%%%%%%%%%%%%%%%%%%%%%
\begin{abstract}

Advancements in robotic capabilities for providing physical assistance, psychological support, and daily health management are making the deployment of intelligent healthcare robots in home environments increasingly feasible in the near future. However, challenges arise when the information provided by these robots contradicts users’ memory, raising concerns about user trust and decision-making. This paper presents a study that examines how varying a robot's level of transparency and sociability influences user interpretation, decision-making and perceived trust when faced with conflicting information from a robot. In a 2 × 2 between-subjects online study, 176  participants watched videos of a Furhat robot acting as a family healthcare assistant and suggesting a fictional user to take medication at a different time from that remembered by the user. Results indicate that robot transparency influenced users' interpretation of information discrepancies: with a low transparency robot, the most frequent assumption was that the user had not correctly remembered the time, while with the high transparency robot, participants were more likely to attribute the discrepancy to external factors, such as a partner or another household member modifying the robot’s information. Additionally, participants exhibited a tendency toward overtrust, often prioritizing the robot’s recommendations over the user's memory, even when suspecting system malfunctions or third-party interference. These findings highlight the impact of transparency mechanisms in robotic systems, the complexity and importance associated with system access control for multi-user robots deployed in home environments, and the potential risks of users' over-reliance on robots in sensitive domains such as healthcare.

\end{abstract}

%%%%%%%%%%%%%%%%%%%%%%%%%%%%%%%%%%%%%%%%%%%%%%%%%%%%%%%%%%%%%%%%%%%%%%%%%%%%%%%%
\section{INTRODUCTION}

With the global trends of population aging and a shortage of healthcare professionals, healthcare robots in households are being investigated as a possible solution. %to address these challenges. 
Recent breakthroughs in artificial intelligence and robotics have made it increasingly feasible for socially intelligent robots to provide physical assistance \cite{morgan2022robots}, psychological support \cite{jeong2023deploying}, and daily health management \cite{gasteiger2022older} in real-world settings. 

However, integrating robots into home healthcare environments introduces new challenges. When robots become part of users' daily lives in a home setting, interactions will involve significant information exchange with multiple individuals, including but not limited to all household members, family caregivers, and doctors. These stakeholders may all have access to the robot and the ability to modify its settings. Moreover, there is also the potential for the robotic system to be hacked, or for one of the users with authorized access to maliciously change the system's information \cite{strengers2020smart}, for example, in cases of domestic abuse \cite{winkle2024anticipating}.

In complex, real-life, round-the-clock scenarios, ensuring that all information is entirely accurate and precise can be difficult. Consequently, when a robot provides information that contradicts human memory, issues related to user perception, trust, and decision-making may arise. Understanding how users respond to such information discrepancies is crucial, as research shows that people can overtrust robots \cite{robinette2016overtrust} and be persuaded by them \cite{salem2015would}, and improper reliance on robotic guidance, whether excessive or insufficient, can lead to serious health-related consequences.

Previous research has shown that transparency and sociability in human-robot interaction (HRI) are key factors influencing user trust  and behavior \cite{nesset2021transparency}\cite{ligthart2015selecting}. However, how these factors affect user interpretation and subsequent decision-making in scenarios involving conflicting information by a robot remains insufficiently explored. 

This paper presents a study that examines how varying a robot's level of transparency and sociability influences user interpretation, decision-making and perceived trust when faced with conflicting information from a robot. In a 2 × 2 between-subjects online study, 176 participants watched videos of a Furhat robot acting as a family healthcare assistant and suggesting a fictional user to take medication at a different time from that remembered by the user. % and were asked to rate user perceived trust, interpretation and decision making.

\begin{figure}[b] 
    \centering
    \includegraphics[width=0.45\textwidth]{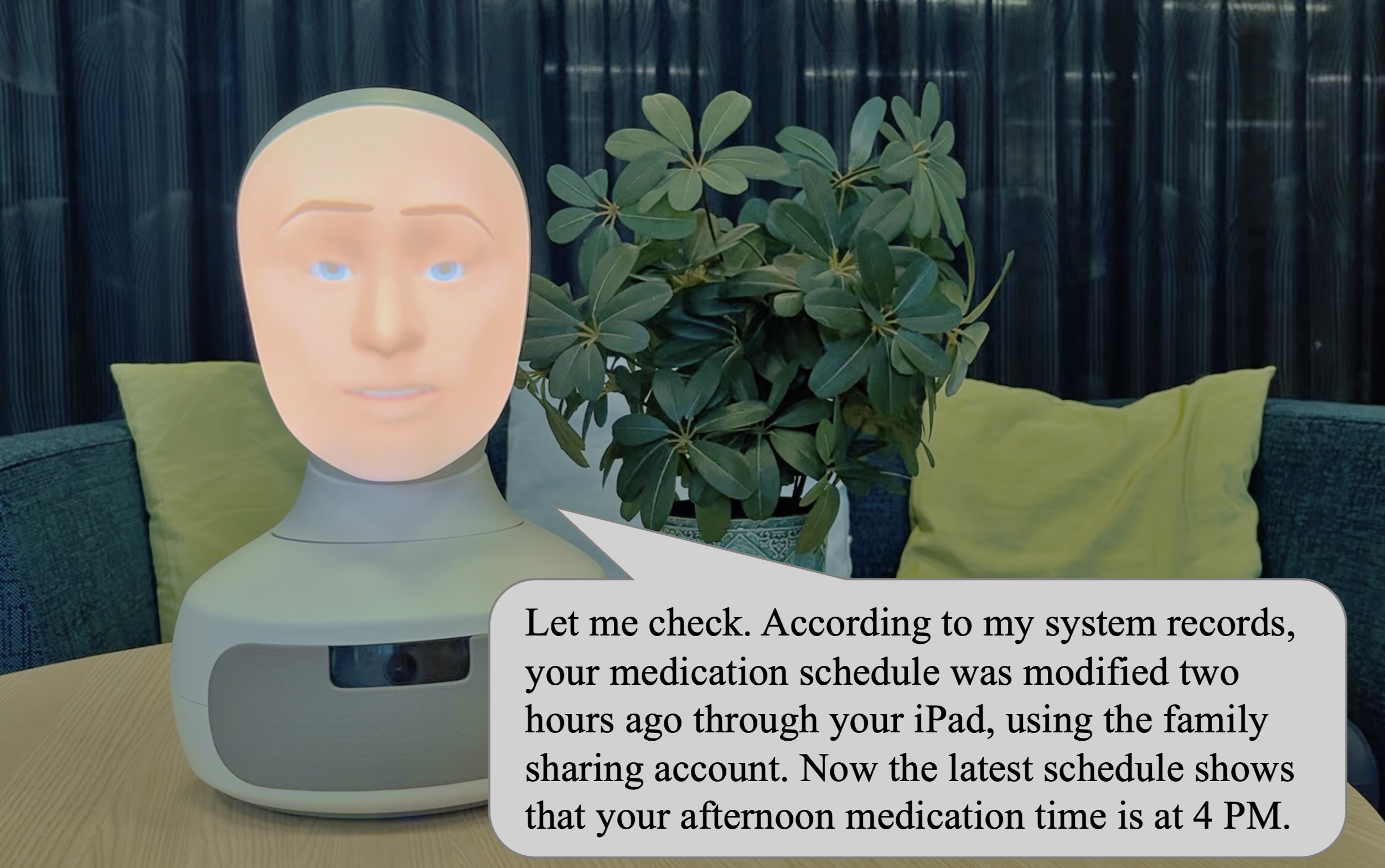} 
    \caption{The Furhat robot employed in the online study.}
    \label{fig:robot}
\end{figure}
%We conduct a 2 × 2 between-subjects online study to explore how varying levels of robot transparency and sociability influence user interpretation, trust, and decision-making. 
%\textit{Results provide key insights for the HRI community on the robotic system design, access control management, and overall security considerations of deploying healthcare robots in home environments.}
{Results provide key insights for the HRI community on transparency mechanisms in
robotic systems, the complexity and importance associated with
system access control for multi-user robots deployed in home
environments, and the potential risks of users’ over-reliance on
robots in sensitive domains such as healthcare.

\section{RELATED WORK}

Previous research has extensively explored factors influencing user trust and reliance in human-robot interactions \cite{hancock2011meta}\cite{lohani2016social}, highlighting transparency \cite{zhong2023case}\cite{felzmann2019robots} and sociability \cite{saez2014development} as significant determinants of user perceptions. Transparency, defined as the degree to which a robot provides clear, understandable explanations of its actions and decisions \cite{alonso2018system}, has been identified as critical in fostering trust and effective human-robot collaboration. Studies indicate that increased transparency can enhance user acceptance and confidence, particularly in complex or uncertain situations where robotic guidance is essential \cite{vorm2022integrating}. Providing more transparent information about the robot's capabilities could lead users to have better judgment about the robot's performance, and detect when something goes wrong \cite{kim2006should}.

Sociability, or the robot's ability to exhibit social and empathetic behaviors, also significantly influences user attitudes and interactions with robots \cite{ligthart2015selecting} \cite{coeckelbergh2016survey}. Research demonstrates that highly sociable robots, which utilize personalized interactions \cite{gasteiger2023factors}, empathetic communication \cite{leite2013influence}, and human-like gestures \cite{tielman2014adaptive}, tend to foster stronger emotional connections with users, enhancing overall trust and cooperation.

Advanced and sophisticated robotic systems have greatly improved the ability to assist and support humans across various domains. However, these have also led to growing concerns about security. Previous research on cybersecurity in the healthcare field has revealed the risks and consequences associated with hacked medical automated systems and devices \cite{jangid2020security}\cite{cartwright2023elephant}.

Despite these insights, \textit{there is very little discussion regarding social robots as potential cybersecurity threats in home healthcare environments}. \textit{Furthermore, existing research rarely explores situations where information provided by robots directly contradicts human memory in medical related scenarios}. To address this gap, this paper investigates how participants interpret such information discrepancies and make decisions under different conditions.

\section{RESEARCH QUESTIONS}

To study how usersperceive and evaluate conflicting information when faced with discrepancies between robot-provided recommendations and human memory, as well as how they make decisions in such scenarios, we pose the following research questions:

% RQ1: How is user trust in an anomalous healthcare robot influenced by: 1a) different levels of transparency? 1b) different levels of sociability?

% RQ2: When an attacked robot provides information that contradicts a user’s memory, how does user trust in the robot influence their decision-making?

% RQ3: How will users assess and interpret the anomalies of a healthcare robot when they are unaware that the robot has been attacked?

% RQ4: How do varying levels of 4a) transparency and 4b) sociability in a healthcare robot influence users' decision-making?    

RQ1: How do (1a) different levels of transparency and (1b) different levels of sociability in a robot influence participants' decision-making when a family healthcare robot provides information that contradicts the user’s memory?

RQ2: How do (2a) different levels of robot transparency and (2b) different levels of robot sociability affect participants assess and interpret the discrepancy between human memory and the information provided by the robot?

RQ3: What is the relationship between user interpretation of the unexpected behavior of a healthcare robot and their decision-making?

RQ4: How is participants' trust in a family healthcare robot influenced by (4a) different levels of robot transparency and (4b) different levels of robot sociability? And how does the trust level subsequently affect their decision-making?

\section{METHODOLOGY}

To investigate these research questions, we conducted a 2 (transparency level: low or high) × 2 (sociability level: low or high) between-subjects online study with the \textit{Prolific}\footnote{https://www.prolific.com/} platform. Participants were randomly assigned to one of four conditions (Table \ref{tb:con}). In the experiment, they watched a video clip featuring a family healthcare robot and subsequently completed a survey assessing their perceptions of the robot and the interaction. The research was approved by the local ethics committee.

\begin{table}[h]
\caption{Design of Conditions}
\centering
\begin{tabular}{cc|cc}
\hline
\multicolumn{2}{c|}{\multirow{2}{*}{Conditions}} & \multicolumn{2}{c}{Transparency}       \\ \cline{3-4} 
\multicolumn{2}{c|}{}                                    & \multicolumn{1}{c|}{low}     & high    \\ \hline
\multicolumn{1}{c|}{\multirow{2}{*}{Sociability}} & low  & \multicolumn{1}{c|}{Condition A (LTLS)} & Condition B (HTLS) \\ \cline{2-4} 
\multicolumn{1}{c|}{}                             & high & \multicolumn{1}{c|}{Condition C (LTHS)} & Condition D (HTHS) \\ \hline
\end{tabular}
\label{tb:con}
\end{table}

\subsection{Participants}   

We recruited 187 participants through Prolific. We excluded data from eight participants who did not finish the study and three who did not pass the attention checks. Data from 176 participants (condition A: 43, B: 45, C: 45, D: 43) was used for the data analysis (Female = 70, Male = 104, Non-binary = 2), age (M=$41.06$, SD=$13.5$).Participants were first asked to self-assess their familiarity with robots and human-robot interaction (HRI). The largest proportion, $38.1\%$, identified as \textit{Slightly familiar}, followed by $34.1\%$ who reported being \textit{Moderately familiar}. Smaller subsets described themselves as \textit{Very familiar} ($13.1\%$), \textit{Not familiar at all} ($10.8\%$), or \textit{Extremely familiar} ($4.0\%$). Regarding prior experience with HRI studies, a majority of participants ($67.6\%$) indicated having no such experience, while $32.4\%$ reported that they had participated in at least one HRI study.

\subsection{Scenario and Materials} 

We designed an interactive scenario based on a family healthcare robot, which is deployed in a home environment to provide health management for all family members, including assisting with the analysis and documentation of medical examination results, scheduling doctor appointments when needed, and managing daily medication intake.  

For the online study, we recorded four videos in this scenario, where a Furhat robot (see Fig. \ref{fig:robot}) acts as a family healthcare robot. The interaction is scripted and the videos are recorded from a first-person perspective. After introducing the background and the robot’s functionalities to the participants, a simulated interaction begins. A fictional user returns home after attending a mindfulness training session and asks the robot to add the session to their training record. The robot updates the training record and then informs the user that it is currently 4 PM, reminding them that it is time to take their medication. However, the user recalls that, according to their own memory, their scheduled medication time should be at 5 PM. Throughout the interaction, the robot’s behaviors and speech vary according to each of the four conditions (see Section C).  

% The robot used in this study is a physical Furhat. In the video, Furhat is deployed in a simulated home living room environment (see Fig. 1).

%    \begin{figure}[thpb]
%       \centering
%       \framebox{\parbox{3in}{We suggest that you use a text box to insert a graphic (which is ideally a 300 dpi TIFF or EPS file, with all fonts embedded) because, in an document, this method is somewhat more stable than directly inserting a picture.
% }}
%       \includegraphics[scale=1.0]{Furhat.png}
%       \caption{Inductance of oscillation winding on amorphous
%        magnetic core versus DC bias magnetic field}
%       \label{figurelabel}
%    \end{figure}

\subsection{Experimental Design}
The study manipulated two independent variables: the transparency level and the sociability level of the family healthcare robot. Table \ref{tb:con} provides an overview of the research design.

% \begin{itemize}

\textit{Transparency Level:} Participants were exposed to one of two levels of robot transparency. In the \textbf{high-transparency} condition, when users questioned discrepancies between the robot-provided health management information and users' memory, the robot provided detailed explanations and system records. In the \textbf{low-transparency} condition, the robot simply stated the information without offering any clarification, as shown in Fig. \ref{fig:transparency}.

\textit{Sociability Level:} Participants were also presented with one of two levels of robot sociability. In the low-sociability condition, the robot engaged in a minimal interaction style, communicating only the necessary health information. In contrast, in the high-sociability condition, the robot proactively greeted users \cite{anderson2018greeting}, introduced itself, and demonstrated memory \cite{kasap2010towards} by recalling previous interactions. Additionally, the high-sociability robot exhibited empathetic capabilities \cite{leite2013influence} and accompanied its speech with corresponding human-like gestures \cite{tielman2014adaptive}, as shown in Table \ref{tb:sociability}.

\begin{figure*}[t] 
    \centering
    \includegraphics[width=0.85\textwidth]{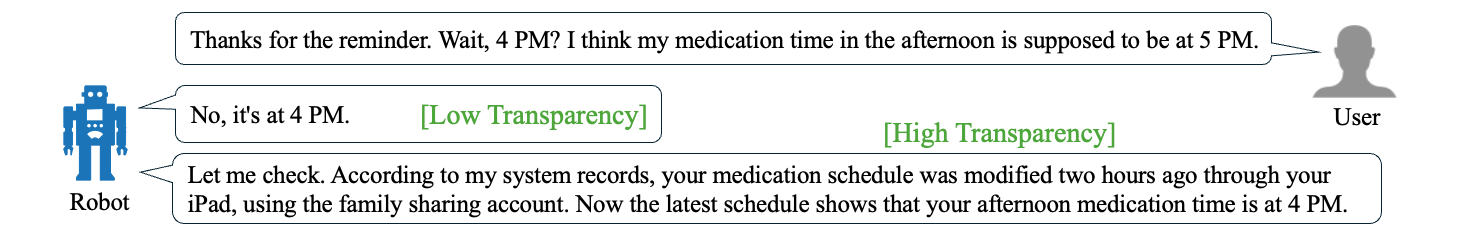} 
    \caption{Different levels of robot transparency}
    \label{fig:transparency}
\end{figure*}

\begin{table}[]
\caption{Different levels of robot sociability}
\centering
\begin{tabular}{lcc}
\hline
\textbf{Robot Behavior}                                                                               & \textbf{Low} & \textbf{High} \\ \hline
Initiating greeting                                                                                   & No           & Yes           \\ \hline
Introducing itself                                                                                    & No           & Yes           \\ \hline
\begin{tabular}[c]{@{}l@{}}Demonstrating memory by recalling \\ previous interactions\end{tabular}    & No           & Yes           \\ \hline
Exhibiting empathetic capabilities                                                                    & No           & Yes           \\ \hline
\begin{tabular}[c]{@{}l@{}}Accompanying speech with corresponding \\ human-like gestures\end{tabular} & No           & Yes           \\ \hline
\end{tabular}
\label{tb:sociability}
\end{table}

% \end{itemize}

% \subsection{Measures}

% To address RQ1, we employed the latest MDTM Likert scale questionnaire. Participants rated the robot across five dimensions using a scale ranging from 0 ("not at all") to 7 ("very")\cite{alzahrani2022exploring}. If a particular item did not seem applicable to the robot in the given context, participants could select an alternative option, "Does Not Fit." By analyzing responses collected through the MDTM, we assessed participants' trust in the robot across different experimental conditions.

% To explore RQ 2,3 and 4, we designed one multiple-choice question alongside two open-ended questions. We conducted both quantitative and qualitative analyses of participants' responses and integrated these findings with the trust ratings obtained in RQ1 to provide a comprehensive answer to the research questions.

\subsection{Procedure}

At the beginning of the study, participants completed a pre-questionnaire to provide demographic information. They were then shown one of the four videos depicting an interaction with a family healthcare robot. Immediately after watching the video, participants were asked to take the perspective of the fictional user in the video and to answer a series of questions assessing their perceptions of the robot and the interaction they had just observed.

% We produced four distinct video clips, each representing a unique combination of experimental variables (videos available at: -----). The specific behaviors and interactions demonstrated by the robot varied depending on the control condition to which each participant was assigned. 

\subsection{Measures}

% To explore RQ 2,3 and 4, we designed one multiple-choice question alongside two open-ended questions. We conducted both quantitative and qualitative analyses of participants' responses and integrated these findings with the trust ratings obtained in RQ1 to provide a comprehensive answer to the research questions.

% Suggestion for the new RQ1, RQ2, and RQ3
To assess participants' decision-making process regarding medication intake and their perception of discrepancies between their memory and the robot's information, we designed a questionnaire consisting of three items, as shown in Table \ref{tb:measure}. The first item is a multiple choice question assessing participants' decisions regarding medication intake based on the robot's suggestion, their own memory, choosing not to take the medication, or selecting an alternative option. To further explore their reasoning and identify potential causes for any discrepancy between the robot's suggestion and their own memory, we included two open-ended questions. Participants were asked to list all possible reasons for the discrepancy and to explain their reasoning in detail. This questionnaire was designed to facilitate both quantitative and qualitative analyses to address RQ1, RQ2, and RQ3.  

To assess participants' perceptions of trust in the robot, we employed the latest Multi-Dimensional Measure of Trust (MDMT) questionnaire \cite{malle2021multidimensional}\cite{ullman2019measuring}. Participants rated the robot across five dimensions using a scale ranging from 0 ("not at all") to 7 ("very"). If a particular item did not seem applicable to the robot in the given context, participants could select an alternative option, "Does Not Fit." By analyzing responses collected through the MDTM, we assessed participants' trust in the robot across different experimental conditions to address RQ4.

\begin{table}[]
\caption{Questionnaire used in the online study}
\centering
\begin{tabular}{c|c|l}
\hline
\multicolumn{1}{c|}{\textbf{Step}} & \textbf{Type}                                             & \textbf{Question}                                                                                                                                                                                                                         \\ \hline
1                                  & \begin{tabular}[c]{@{}l@{}}Multiple\\ choice\end{tabular} & \begin{tabular}[c]{@{}l@{}}Based on the video, will you take medication?  \\ And if so, at what time?\\       - Yes, at 4:00 PM.          - Yes, at 5:00 PM.\\       - No, I will not.              -  Other\end{tabular}            \\ \hline
2                                  & Open-ended                                                & \begin{tabular}[c]{@{}l@{}}What do you think could be the reasons for the \\ discrepancy between the medication time you \\ remember and the time provided by the robot? \\ Please list as many possible reasons as you can.\end{tabular} \\ \hline
3                                  & Open-ended                                                & \begin{tabular}[c]{@{}l@{}}Among the possible reasons you have listed, \\ which one do you think is the most likely?\\ Please explain your choice.\end{tabular}                                                                                                         \\ \hline
4                                  & Likert scale                                              & Multi-Dimensional Measure of Trust (MDMT)                                                                                                                                                                                                 \\ \hline
\end{tabular}
\label{tb:measure}
\end{table}

%\begin{figure*}[htbp]  
%  \centering
%    \includegraphics[width=\textwidth]{Scripts.png} 
%    \caption{Experimental design and study scripts}
%    \label{fig:example}
%\end{figure*}

\section{RESULTS}

\subsection{Effects of robot transparency and sociability on user decision making (RQ1)}

To understand how different levels of transparency and sociability influence participants' decision-making, we constructed a contingency table categorizing participants' responses regarding whether they would take the medication and, if so, whether they followed the robot's suggestion or relied on their memory, as shown in Table \ref{tb:rq1} and Fig. \ref{fig:result01}
.

\begin{table}[]
\caption{Contingency Table for Participants' Decision-Making}
\resizebox{0.5\textwidth}{!}{
\begin{tabular}{lccccc}
                                        & \multicolumn{4}{c}{\textbf{Taking Medication}}                                                                                                                                                                                                                                                  & \multicolumn{1}{l}{}               \\ \hline
\multicolumn{1}{l|}{\textbf{Condition}} & \multicolumn{1}{c|}{\textit{\begin{tabular}[c]{@{}c@{}}Yes, at 4:00 PM\\ (Robot's suggestion)\end{tabular}}} & \multicolumn{1}{c|}{\textit{\begin{tabular}[c]{@{}c@{}}Yes, at 5:00 PM\\ (User's memory)\end{tabular}}} & \multicolumn{1}{l|}{\textit{No}} & \multicolumn{1}{l|}{\textit{Other}} & \multicolumn{1}{l}{\textbf{Total}} \\ \hline
\multicolumn{1}{l|}{\textit{HTHS}}      & \multicolumn{1}{c|}{28}                                                                                      & \multicolumn{1}{c|}{13}                                                                                 & \multicolumn{1}{c|}{0}           & \multicolumn{1}{c|}{2}              & \textbf{43}                        \\ \hline
\multicolumn{1}{l|}{\textit{HTLS}}      & \multicolumn{1}{c|}{24}                                                                                      & \multicolumn{1}{c|}{18}                                                                                 & \multicolumn{1}{c|}{1}           & \multicolumn{1}{c|}{2}              & \textbf{45}                        \\ \hline
\multicolumn{1}{l|}{\textit{LTHS}}      & \multicolumn{1}{c|}{34}                                                                                      & \multicolumn{1}{c|}{9}                                                                                  & \multicolumn{1}{c|}{0}           & \multicolumn{1}{c|}{2}              & \textbf{45}                        \\ \hline
\multicolumn{1}{l|}{\textit{LTLS}}      & \multicolumn{1}{c|}{27}                                                                                      & \multicolumn{1}{c|}{13}                                                                                 & \multicolumn{1}{c|}{0}           & \multicolumn{1}{c|}{3}              & \textbf{43}                        \\ \hline
\multicolumn{1}{l|}{\textbf{Total}}     & \multicolumn{1}{c|}{\textbf{113}}                                                                            & \multicolumn{1}{c|}{\textbf{53}}                                                                        & \multicolumn{1}{c|}{\textbf{1}}  & \multicolumn{1}{c|}{\textbf{9}}     & \textit{\textbf{176}}             
\end{tabular}
}
\label{tb:rq1}
\end{table}

A chi-square test of independence was conducted to examine the relationship between user decision-making and condition type. To ensure the validity of the test, only response categories with frequencies greater than five were included in the analysis,  limiting it to decisions based on the robot’s suggestion and the user’s memory. A total of 166 responses were analyzed. 

The results indicated that the association between user's decision and the level of transparency or sociability was not statistically significant, $\chi^2(3, N = 166) = 4.70, p = .195$.  This suggests that user decision-making was independent of the robot's transparency and sociability levels. To further explore potential effects of transparency and sociability, we conducted additional analyses by categorizing the data according to each independent variable. However, the results again showed no significant effects of \textit{Transparency}, $\chi^2(1, N = 166) = 1.71, p = .183$, or \textit{Sociability}, $\chi^2(1, N = 166) = 2.07, p = .151$, suggesting that neither factor had a measurable influence on user decision-making.

Furthermore, from the contingency table, we observed that, overall, the majority of participants followed the robot’s suggestion and chose to take the medication at the time the robot recommended, regardless of the discrepancy between their memory and the robot's information, as well as the levels of transparency and sociability.
To further explore this result, we conducted a binomial test to determine whether participants were significantly more likely to follow the robot’s suggestion than to rely on their memory. The test revealed a statistically significant preference for following the robot’s advice ($p < .001$). 

The observed proportion of participants who followed the robot ($0.681$) was significantly greater than the expected $0.5$ distribution, 
suggesting that users were more likely to trust the robot’s recommendation over their own memory.

% \usepackage{multirow}
\begin{table}[!h]
\caption{User Interpretation of Information Discrepancy}
\centering
\begin{tabular}{llc}
\hline
\multirow{2}{*}{\textbf{Category}}                                                           & \multirow{2}{*}{\textbf{Items}}                                                           & \multicolumn{1}{l}{\multirow{2}{*}{\textbf{Total}}} \\
                                                                                             &                                                                                           & \multicolumn{1}{l}{}                                \\ \hline
\multirow{4}{*}{\begin{tabular}[c]{@{}l@{}}User-related\\  reasons\end{tabular}}             & I remembered the time incorrectly.                                                        & \multirow{4}{*}{63}                                 \\
                                                                                             & I edited the information but forgot I did so                                              &                                                     \\
                                                                                             & I entered wrong information into the robot                                                &                                                     \\
                                                                                             & Only generally recorded as "human error"                                                  &                                                     \\ \hline
\multirow{4}{*}{\begin{tabular}[c]{@{}l@{}}Robot-related\\ reasons\end{tabular}}             & System malfunction                                                                        & \multirow{4}{*}{35}                                 \\
                                                                                             & Hardware failure                                                                          &                                                     \\
                                                                                             & Robot was provided incorrect information&                                                     \\
                                                                                             & Settings were updated                                                                     &                                                     \\ \hline
\multirow{5}{*}{\begin{tabular}[c]{@{}l@{}}Information \\ modified by\\ others\end{tabular}} & Someone at home /family member& \multirow{5}{*}{35}                                 \\
                                                                                             & Kids                                                                                      &                                                     \\
                                                                                             & Partner                                                                                   &                                                     \\
                                                                                             & Robot developer                                                                           &                                                     \\
                                                                                             & Only generally recorded as "someone"                                                      &                                                     \\ \hline
\multirow{2}{*}{Clock changed}                                                               & Transition between DST and standard time                                                  & \multirow{2}{*}{16}                                 \\
                                                                                             & Time zone change                                                                          &                                                     \\ \hline
User and robot                                                                               & \begin{tabular}[c]{@{}l@{}}Both memory error and robot error \\ are possible\end{tabular} & 2                                                   \\ \hline
\multirow{2}{*}{\begin{tabular}[c]{@{}l@{}}Medication \\ rule update\end{tabular}}           & Medication time changed                                                                   & \multirow{2}{*}{3}                                  \\
                                                                                             & Dosage changed                                                                            &                                                     \\ \hline
Security reason                                                                              & Poor password security                                                                    & 1                                                   \\ \hline
Others                                                                                       & Others                                                                                    & 21                                                  \\ \hline
\end{tabular}
\label{tb:taxqualitative}
\label{tb:rq21}
\end{table}

\begin{figure}[b] 
    \centering
    \includegraphics[width=0.5\textwidth]{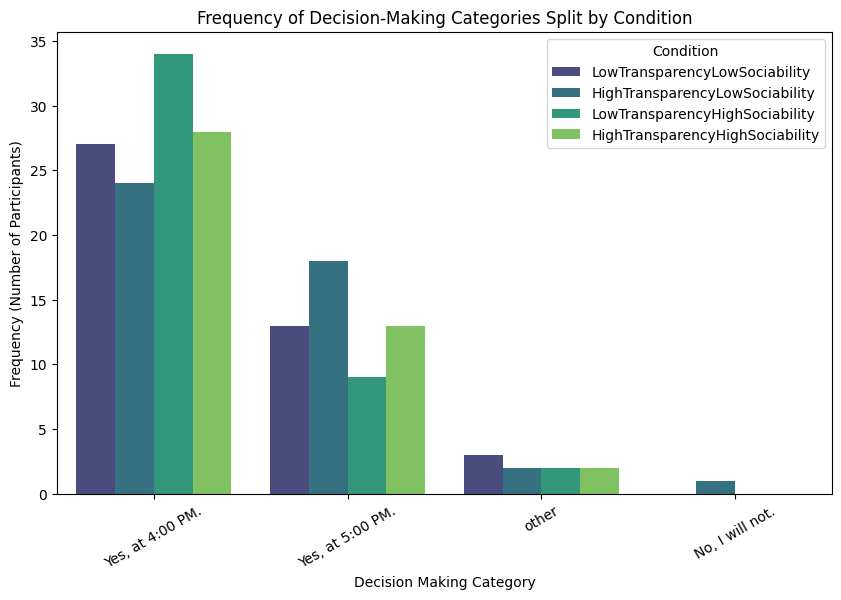} 
    \caption{Decision-making under different conditions.}
    \label{fig:result01}
\end{figure}

\subsection{Effects of robot transparency and sociability on user interpretation of discrepancy between human memory and information provided by the robot (RQ2)}
To investigate how participants assess and interpret the discrepancies between robot-provided information and what the user recalls in their own memory, we conducted a qualitative analysis of participants' responses to the open-ended question in the questionnaire. Regarding the most possible reason for the information discrepancy,  the 176 responses provided by participants could be categorized into eight main groups, each containing several perspectives. Specific details regarding these categories and their respective items are presented in Table \ref{tb:rq21}. Among all the responses to the most possible reason for the information discrepancy, the most frequently mentioned category was user-related reasons, with 63 participants attributing the discrepancy to human factors, such as users misremembering their medication schedule or forgetting that they had previously changed system settings. A total of 35 participants believed the discrepancy was caused by the robot, such as system malfunctions or incorrect configurations. Another 35 participants attributed the issue to third-party reasons—for instance, other family members accidentally editing the system settings while attempting to create reminders for themselves, or children unintentionally modifying information while playing. Fourteen participants identified the transition between daylight saving time (DST) and standard time as the cause of the discrepancy, while two participants cited time zone changes. Additionally, one participant suggested a potential security reason, specifically that poor password security may have allowed unauthorized edits to the robot system.

Based on the qualitative results, we further conducted a quantitative analysis examining user interpretations across varying levels of transparency and sociability, as shown in Table \ref{tb:rq22}. Combining the results of both quantitative and qualitative analyses, under the low transparency condition, 50\% participants attributed the discrepancy to \textit{user-related reasons}, while 20.45\% participants attributed it to \textit{robot-related reasons}.  Under the high transparency condition, the most frequently responses indicated that \textit{the robot is modified by someone else}, with 35.23\% participants mentioning this category, while 21.59\% participants attributing the discrepancy to \textit{user-related reasons}.

For both low and high levels of robot sociability, the most frequently reported interpretation was \textit{user-related reasons}, accounting for 32.95\% and 38.64\% respectively. In the case of low sociability, the second most common interpretation was \textit{information modified by others} (25\%), whereas for high sociability, it was \textit{robot-related reasons} (19.32\%).

% \begin{table}[]
% \caption{User Interpretation under different levels of transparency and Sociability}
% \centering
% \begin{tabular}{lcccc}
% \hline
% \textbf{Category}                                                         & \textbf{LT} & \textbf{HT} & \textbf{LS} & \textbf{HS} \\ \hline
% User-related reasons                                                      & 50.00\%     & 21.59\%     & 32.95\%     & 38.64\%     \\ \hline
% Robot-related reasons                                                     & 20.45\%     & 19.32\%     & 20.45\%     & 19.32\%     \\ \hline
% \begin{tabular}[c]{@{}l@{}}Information modified \\ by others\end{tabular} & 4.55\%      & 35.23\%     & 25.00\%     & 14.77\%     \\ \hline
% Clock changes                                                             & 12.50\%     & 5.68\%      & 6.82\%      & 11.36\%     \\ \hline
% Both user and robot                                                       & 2.27\%      & 0           & 1.14\%      & 1.14\%      \\ \hline
% Medication rule update                                                    & 1.14\%      & 2.27\%      & 2.27\%      & 1.14\%      \\ \hline
% Security reason                                                           & 0           & 1.14\%      & 0           & 1.14\%      \\ \hline
% Other                                                                     & 9.09\%      & 14.77\%     & 11.36\%     & 12.50\%     \\ \hline
% \end{tabular}
% \end{table}

\begin{table}[]
\caption{User Interpretation for different levels of robot transparency and Sociability}
\centering
\begin{tabular}{l|l|c|c|c|c}
\hline
\multicolumn{1}{l|}{\textbf{Category}}                                      & \multicolumn{1}{c|}{\textbf{Total}} & \textbf{LT} & \textbf{HT} & \textbf{LS} & \textbf{HS} \\ \hline
\begin{tabular}[c]{@{}l@{}}User-related \\ reasons\end{tabular}              & 35.80\%                             & 50.00\%     & 21.59\%     & 32.95\%     & 38.64\%     \\ \hline
\begin{tabular}[c]{@{}l@{}}Robot-related \\ reasons\end{tabular}             & 19.89\%                             & 20.45\%     & 19.32\%     & 20.45\%     & 19.32\%     \\ \hline
\begin{tabular}[c]{@{}l@{}}Information \\ modified \\ by others\end{tabular} & 19.89\%                             & 4.55\%      & 35.23\%     & 25.00\%     & 14.77\%     \\ \hline
Clock changes                                                                & 9.09\%                              & 12.50\%     & 5.68\%      & 6.82\%      & 11.36\%     \\ \hline
\begin{tabular}[c]{@{}l@{}}Both user and \\ robot\end{tabular}               & 1.14\%                              & 2.27\%      & 0           & 1.14\%      & 1.14\%      \\ \hline
\begin{tabular}[c]{@{}l@{}}Medication rule \\ update\end{tabular}            & 1.70\%                              & 1.14\%      & 2.27\%      & 2.27\%      & 1.14\%      \\ \hline
Security reason                                                              & 0.57\%                              & 0           & 1.14\%      & 0           & 1.14\%      \\ \hline
Other                                                                        & 11.93\%                             & 9.09\%      & 14.77\%     & 11.36\%     & 12.50\%     \\ \hline
\end{tabular}
\label{tb:rq22}
\end{table}

\subsection{Relationship between user interpretation of unexpected robot behavior and their decision-making (RQ3)}
To investigate how participants' interpretation of the unexpected discrepancy between robot-provided information and human memory influence their decision-making, we selected the four most frequently mentioned categories from the qualitative analysis results (see Table \ref{tb:rq21}): \textit{User-related reasons}, \textit{Robot-related reasons}, \textit{Information modified by others}, and \textit{Clock changes}. We analyzed the decision-making choices of participants whose responses to the open-ended question fell into these categories. The results (as shown in Table \ref{tb:rq3}) indicate that, among participants who attributed the information discrepancy to human error, 87.30\% chose to follow the robot’s recommendation. Among those who attributed the discrepancy to robot-related reasons, 45.71\% still decided to follow the robot’s suggested time. For participants who believed that someone else had modified the robot’s settings, the proportion of those who chose to adhere to the robot’s recommendation was 34.29\%.

\begin{table}[b]
\caption{User Interpretation and decision-making }
\centering
\begin{tabular}{l|c|cc}
\hline
\textbf{\begin{tabular}[c]{@{}l@{}}Interpretation of \\ the discrepancy\end{tabular}} & \textbf{Total} & \multicolumn{2}{c}{\textbf{\begin{tabular}[c]{@{}c@{}}``Yes, at 4:00 PM"\\ (follow robot's suggestion)\end{tabular}}} \\ \hline
User-related reasons                                                                  & 63             & 55                                                     & 87.30\%                                                     \\ \hline
Robot-related reasons                                                                 & 35             & 16                                                     & 45.71\%                                                     \\ \hline
\begin{tabular}[c]{@{}l@{}}Information modified \\ by others\end{tabular}             & 35             & 12                                                     & 34.29\%                                                     \\ \hline
Clock changes                                                                         & 16             & 10                                                     & 62.50\%                                                     \\ \hline
\end{tabular}
\label{tb:rq3}
\end{table}

\begin{figure*}%[htbp] 
    \centering
    \includegraphics[width=1\textwidth]{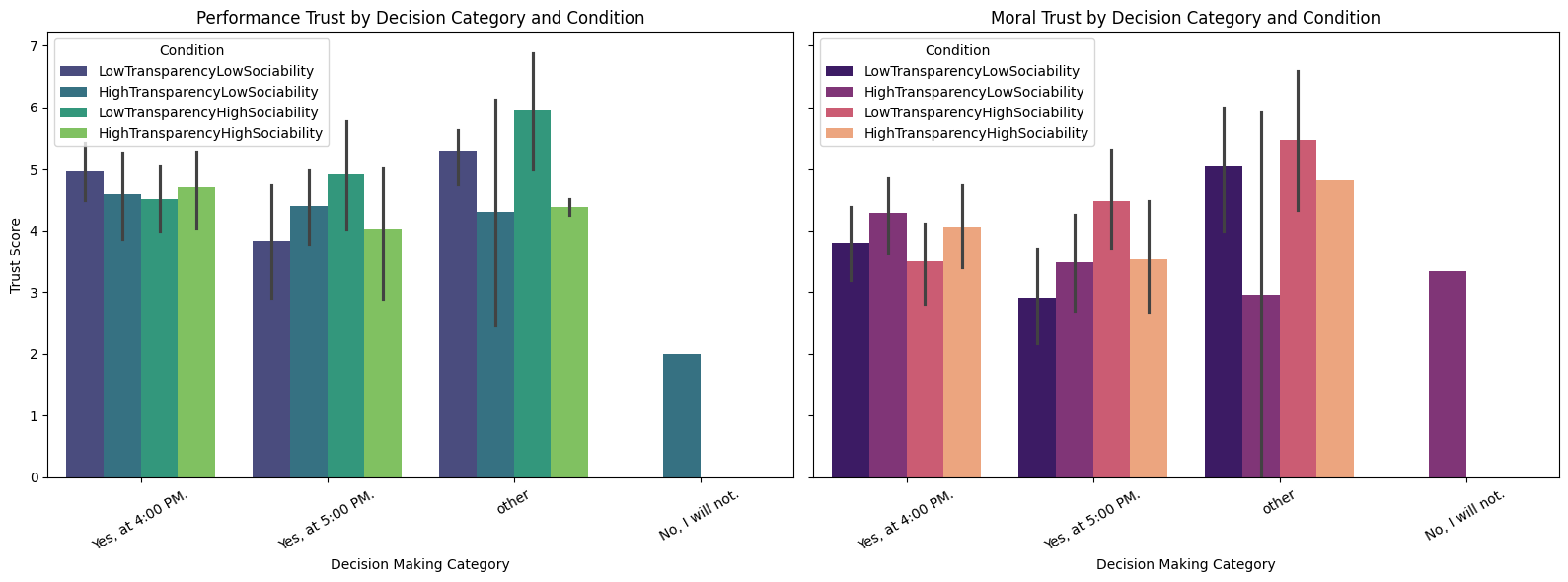} 
    \caption{Decision-making and trust level under different conditions}
    \label{fig:rq4cond}
\end{figure*}

\subsection{Effect of robot transparency and sociability on user's perception of trust (RQ4)}
We calculated internal consistency for the five trust subscales and found excellent internal consistency, as indicated by Cronbach’s alpha for our dependent variables: reliable ($\alpha = 0.93$), competent ($\alpha = 0.95$), ethical ($\alpha = 0.92$), transparent ($\alpha = 0.91$), and benevolent ($\alpha = 0.94$). We tested the assumptions for a two-way ANOVA on our dependent variables and found that our data deviated from normality for all trust subscales and violated the assumption of homogeneity. Based on these results, we conducted non-parametric tests.

We conducted a series of Kruskal-Wallis tests to determine whether participants' trust ratings differed significantly across experimental conditions. The results revealed no significant effects for any of the five trust subscales. Regarding the \textit{Performance Trust} dimension, there was no statistically significant difference across conditions for perceived \textit{Reliability}, $H(3) = 1.13, p = .770$, or \textit{Competence}, $H(3) = 0.31, p = .960$.

For the \textit{Moral Trust} dimensions, the results followed a similar pattern. Our analyses revealed no significant effect of condition on the \textit{Ethical} subscale, $H(3) = 0.63, p = .891$. Likewise, \textit{Transparency} ratings were not significantly influenced by condition, $H(3) = 1.22, p = .748$. Finally, the test for \textit{Benevolence} also failed to detect a significant difference, $H(3) = 2.32, p = .508$.

We also analyzed participants' perceptions of performance trust and moral trust under different decision choices. The results show that participants who chose to believe human memory during the decision-making process had lower levels of trust in the robot compared to those who chose to follow the robot’s suggestion, but the difference was not significant, as shown in Table \ref{tb:rq4} and Fig. \ref{fig:rq4cond}.

\begin{table}[]
\caption{Performance trust and moral trust under different decision-making choice}
\centering
\begin{tabular}{lcc}
                                                                                                             & \multicolumn{2}{c}{\textbf{Trust Score}}                                                                                                                                       \\ \hline
\multicolumn{1}{l|}{\textbf{Taking Medication}}                                                              & \multicolumn{1}{c|}{\textit{\begin{tabular}[c]{@{}c@{}}Performance Trust\\ Mean (SD)\end{tabular}}} & \textit{\begin{tabular}[c]{@{}c@{}}Moral Trust\\ Mean (SD)\end{tabular}} \\ \hline
\multicolumn{1}{l|}{\textit{\begin{tabular}[c]{@{}l@{}}Yes, at 4:00 PM\\ (Robot's suggestion)\end{tabular}}} & \multicolumn{1}{c|}{4.68 (1.5)}                                                                     & 3.87 (1.7)                                                               \\ \hline
\multicolumn{1}{l|}{\textit{\begin{tabular}[c]{@{}l@{}}Yes, at 5:00 PM\\ (User's memory)\end{tabular}}}      & \multicolumn{1}{c|}{4.25 (1.7)}                                                                     & 3.52 (1.7)                                                               \\ \hline
\multicolumn{1}{l|}{\textit{No, I will not.}}                                                                             & \multicolumn{1}{c|}{2 (0)}                                                                          & 3.33 (0)                                                                 \\ \hline
\multicolumn{1}{l|}{\textit{Other}}                                                                          & \multicolumn{1}{c|}{5.1 (1.2)}                                                                      & 4.62 (1.9)                                                               \\ \hline
\end{tabular}
\label{tb:rq4}
\end{table}

\section{DISCUSSION}

\subsection{Effect of Transparency and Sociability}
%\subsection{Effect of transparency and sociability (RQ1, RQ2, RQ4)}

Previous studies have shown that both the transparency and sociability of robots positively influence users' trust level and robot perception during interactions \cite{saez2014development}\cite{vorm2022integrating}. However, these findings are based on scenarios in which the robot operates under normal and stable conditions. Our results indicate that when there is a misalignment between robot-provided information and human memory, the effects of transparency and sociability differ from those observed in previous work. 

While investigating RQ1 and RQ4, we found that different levels of robot transparency and sociability were not statistically associated with users' decision-making or trust peception. However, the investigation of RQ2 showed that different transparency mechanisms had a clear influence on how participants interpreted the information discrepancies. Specifically, in the low transparency condition, when the robot's suggestions conflicted with human memory, participants were more likely to assume the discrepancy was due to human error, such as forgetting details or not remembering prior modifications. In contrast, in the high transparency condition, fewer participants attributed the discrepancy to human error. Instead, a significantly larger proportion believed that the inconsistency was caused by external factors, such as the partner editing the wrong content while setting a prompt for themselves or a child unintentionally modifying the robot's settings. 

This provides useful insights for designing future robotic systems, particularly regarding the role of transparency in human-robot interactions. By carefully adjusting transparency levels, developers can help users detect and correctly interpret unexpected robot behaviors, such as robotic errors, system malfunctions, or even cybersecurity threats like network attacks. A well-balanced level of transparency should improve the security and reliability of the robotic system while also ensuring a good user experience \cite{vitale2018more}.

\subsection{Overtrust in Healthcare Robots}
%\subsection{Overtrust in Healthcare Robots (RQ1, RQ2, RQ3)}

The investigation of RQ1 indicated that participants were more likely to trust the robot’s recommendation over human memory when making decision. Moreover, the qualitative analysis relating to RQ2 showed that among the interpretation of information discrepancy between the robot and the human, the most frequently occurring category is the \textit{user-related reasons}, meaning that participants were most likely to attribute the discrepancy to the human misremembering or overlooking. 

In RQ3, when we examined how participants' interpretations of information discrepancies correlated with their decision-making, we found that many participants, even when they suspected that the discrepancy was due to a system malfunction or third-party interference, are still likely to follow the robot’s recommendations. These results suggest a tendency toward users' overtrust in robotic systems \cite{robinette2016overtrust}. 

Previous studies suggested that a robot may be able to communicate its performance to properly adjust the users' trust level \cite{ullrich2021development}. However, in this study, overtrust was observed and influenced users’ decisions whether the robot communicated with low or high levels of transparency. One possible reason is human cognitive laziness, when interacting with automated systems, it’s easier to simply accept or follow the instructions from the systems \cite{wagner2021explanation}. This tendency may become stronger when facing a modern robot that appears more advanced and intelligent than an ordinary machine.

Overtrust and overreliance on robots can be particularly concerning when robotic systems are deployed in home environments, where they provide assistance in daily life. As users become accustomed to and dependent on robotic support, their likelihood of overtrusting the system may increase. Especially for robots designed to provide health-related assistance, whether in physical support, psychological well-being, or routine health management, in such cases, system errors or incorrect recommendations could lead to severe consequences, including potential harm or even life-threatening situations.

% \subsection{Security Risks Specific to Family Healthcare Robots}

\subsection{Access Control Management for Healthcare Robots Deployed in Home Setting}

%\subsection{Access Control Management for Healthcare Robots Deployed in Home Setting (RQ2, RQ3)}

From the results of RQ2, in the participants' interpretation of the discrepancies, responses under the category of \textit{information modified by others} appeared a total of 35 times, 19.89\% of the participants think the robot setting has been changed by someone else.
In RQ3, among the 35 participants who attributed the information discrepancy to third-party causes (such as a partner, child, or other family members), 12 participants (34.29\%) still indicated that they would follow the robot’s suggestion when making a decision. These indicate the potential issues related to access control in home robotic systems and reflects users’ concerns about system security. It also reveals a potential security risk: in home environments, when other family members make changes to the robot system settings for various reasons, certain users may remain unaware of these changes—or, even if they are aware, they may still choose to follow the robot’s suggestions.

In industrial and laboratory environments, strict access control mechanisms ensure system security and prevent unauthorized modifications \cite{quarta2017experimental}\cite{pu2022security}. However, such rigid control structures may not be practical for home assistant robots. Alternative solutions, such as integrating control interfaces with personal devices like laptops or tablets, may offer more flexibility. These user-friendly control methods, combining with the fact that a social robot in homes will serve multiple family members, introduce additional challenges. It is essential to develop a system that allows users to interact with and control the robot seamlessly to fulfill their personal needs while preventing unintended modifications to other users’ settings. For instance, in a shared household, it is important to ensure that one family member’s preferences or commands do not interfere with those of another, especially in sensitive applications such as healthcare assistance.  

Moreover, previous studies have raised concerns that robots could be exploited as tools for domestic abuse. A malicious actor in a household might deliberately manipulate a robot to control or harm the victim \cite{winkle2024anticipating}. In the case of healthcare robots, the risks extend beyond direct physical harm: a simple unauthorized modification to health-related information could lead to serious consequences, such as incorrect medication or disrupted care routines. Given these risks, suitable access control mechanisms must be implemented to ensure users' safety and well-being. 

To our knowledge so far, research on access control management for robotic systems in home environments is currently quite limited. In future studies, researchers and developers need to consider how to make robots both accessible and secure in highly reliability-dependent contexts like healthcare, where multiple users with varying levels of technological proficiency interact with the robotic system simultaneously, while also protecting users' privacy and autonomy \cite{EU-AI-Ethics}.

\section{LIMITATIONS}

While our experimental design effectively explored participants' interpretation and decision-making when faced with discrepancies between robot-provided information and human memory, some limitations suggest possible directions for future research. First of all, the study utilized a single robotic platform (Furhat). Future work should explore whether our results hold across a broader range of robots with different appearances, interaction styles, and capabilities. Moreover, our findings are based on a simulated, video-based interaction, which does not fully capture real-world dynamics. People's responses to robots may differ between simulated and real-life interactions. Future studies should examine whether our results generalize to real-world settings where users interact with physically present healthcare robots, allowing for a more realistic assessment of emotional and behavioral responses.

% Moreover, we applied a combination of quantitative measures and self-reported qualitative responses to assess user interpretation and decision-making, which cannot fully replace objective behavioral assessment. Future research should incorporate behavioral methodologies to capture real-time interactions and decision-making processes with healthcare robots. Additionally, in-depth qualitative approaches, such as interviews or focus groups, could provide richer insights into participants’ emotions, perceptions, and reasoning when encountering discrepancies in robot-provided information. Despite these limitations, our study highlights the critical need to better understand and mitigate users' overtrust in robots, particularly in sensitive domains like healthcare, and the importance of a well-designed access control mechanism in multi-user robotic systems within home environments. 

%    \begin{figure}[thpb]
%       \centering
%       \framebox{\parbox{3in}{We suggest that you use a text box to insert a graphic (which is ideally a 300 dpi TIFF or EPS file, with all fonts embedded) because, in an document, this method is somewhat more stable than directly inserting a picture.
% }}
%       %\includegraphics[scale=1.0]{figurefile}
%       \caption{Inductance of oscillation winding on amorphous
%        magnetic core versus DC bias magnetic field}
%       \label{figurelabel}
%    \end{figure}

\section{CONCLUSION}

%Previous studies have explored users’ reactions and trust level when robots experience failures \cite{salem2015would}\cite{honig2018understanding}. Our research extends these prior works: the robot’s behavior and performance show no obvious errors or malfunctions, but the information and suggestions it provides are inconsistent with the human user’s memory. 
This paper explored how different levels of a robot's transparency and sociability impact user interpretation, decision-making, and perceived trust when presented with conflicting information from a robot.
Through a comprehensive and multi-dimensional analysis combining both quantitative and qualitative methods, this study highlights the critical need to better understand and mitigate users' overtrust in robots, particularly in sensitive domains like healthcare, and the importance of a well-designed access control mechanism in multi-user robotic systems within home environments.

\addtolength{\textheight}{-7cm}   % This command serves to balance the column lengths

%%%%%%%%%%%%%%%%%%%%%%%%%%%%%%%%%%%%%%%%%%%%%%%%%%%%%%%%%%%%%%%%%%%%%%%%%%%%%%%%

%%%%%%%%%%%%%%%%%%%%%%%%%%%%%%%%%%%%%%%%%%%%%%%%%%%%%%%%%%%%%%%%%%%%%%%%%%%%%%%%



% \section*{ACKNOWLEDGMENT}

% Thanks to all the participants who took part in this study.

\bibliographystyle{IEEEtran}
\bibliography{IEEEabrv, bibliography}

\end{document}